\documentclass[]{ceurart}

\usepackage{todonotes}

\begin{document}

\copyrightyear{2026}
\copyrightclause{Copyright for this paper by its authors.
  Use permitted under Creative Commons License Attribution 4.0
  International (CC BY 4.0).}

\conference{}

\title{Extending the MM-AR Metamodeling Platform with Process Mining Capabilities}

\author[1]{Geraldine Hürzeler}[
email=geraldine.huerzeler@unifr.ch,
url=https://github.com/shuerzeler,
]
\address[1]{University of Fribourg, Bd de Pérolles 90, 1700 Fribourg, Switzerland}

\author[1]{Hans-Georg Fill}[
orcid=0000-0001-5076-5341,
email=hans-georg.fill@unifr.ch
]

\author[1]{Daniel Borcard}[
orcid=0000-0001-7211-4793,
email=daniel.borcard@unifr.ch
]

\fntext[1]{These authors contributed equally.}

\begin{abstract}
Process mining derives process models from event data recorded by information systems, but its results are usually presented as two-dimensional visualizations that are separate from the environments in which enterprise models are created and maintained. Metamodeling platforms for extended reality (XR), such as MM-AR, allow users to define modeling languages and to display models in 2D and 3D, but they do not provide access to event data or mining algorithms so far.
In this paper we therefore present a first proof-of-concept for showing how an XR-based metamodeling platform can be extended with process discovery capabilities without changes to the platform core through a separate Python service based on the open source PM4Py process mining library. 
We demonstrate the approach using three common discovery algorithms for obtaining Petri nets and BPMN models based on CSV or XES event logs. A technical evaluation using four public event logs shows that the element counts of the created model instances match those of the PM4Py models in all 16 configurations. The integration adds between about 8 and 53 seconds per run, and this overhead grows with the size of the discovered model rather than the size of the event log. Although the prototype does not yet use the third dimension to encode process information, it established the foundations for studying such approaches.
\end{abstract}


\maketitle
\section{Introduction}
\label{sec:introduction}
Process mining derives process models from the event data recorded by information systems~\cite{AalstEtAl2012Manifesto}. For enterprise modeling, such discovered models complement models elicited from stakeholders with models based on recorded behaviour. 
At the same time, platforms for conceptual modeling in extended reality (XR), such as MM-AR~\cite{muffMMARWebbasedOpensource2026}, allow users to define modeling languages and interact with the resulting models in 2D and 3D. Previous approaches that bring process mining into virtual reality created dedicated clients for particular languages~\cite{wetzelEntwicklungVRUmgebungZur2022,oberhauserVRProcessMineImmersiveProcess2022}. However, the integration of process mining into an XR metamodeling platform that supports multiple modeling languages has so far not been accomplished.

During discussions with the developers of the MM-AR metamodeling platform we derived the following requirements for such an integration.
(R1)~\emph{Event log upload:} users must be able to upload an event log, since metamodeling platforms have no notion of event data and thus no way to pass such data to an analysis. (R2)~\emph{Mined models as model instances:} discovery results must be transformed into model instances that conform to the platform's meta\textsuperscript{2}-model. (R3)~\emph{Loose coupling of the process mining extension:} as the platform is under continuous, independent development, a tightly coupled integration would be difficult to maintain. (R4)~\emph{Usability for non-experts:} the approach should be easily usable without expert knowledge in process mining.

In the following we will present our approach for meeting these requirements in the form of a proof-of-concept implementation and conduct a first evaluation.

The remainder of the paper is structured as follows.
Section~\ref{sec:background} introduces process mining and MM-AR, and Section~\ref{sec:related-work} discusses related work. Section~\ref{sec:tool-selection} justifies the choice of PM4Py. Sections~\ref{sec:architecture} and~\ref{sec:implementation} present the architecture and the implementation, and Section~\ref{sec:evaluation} demonstrates and evaluates the prototype. Section~\ref{sec:limitations} discusses limitations, and Section~\ref{sec:conclusion} concludes the paper.

\section{Background}
\label{sec:background}
In this section we briefly present background information on process mining and the MM-AR platform to familiarize readers with the most important concepts of these approaches.

\subsection{Process Mining}

Driven by the growing volume of event data recorded by information systems~\cite{Aalst2012ProcessMiningCACM}, process mining uses this data to discover, monitor and improve the processes actually executed in an organization~\cite{AalstEtAl2012Manifesto}. For enterprise modeling, it complements models created from stakeholders with models derived from recorded behavior.
The input is an event log, in which each event refers to a case (one process execution, e.g.\ a purchase order), an activity and usually a timestamp~\cite{bookProcessMining}. These are the three attributes users map when uploading a CSV file to the prototype. Besides generic CSV files, logs are
stored in dedicated formats: the case-centric XES and the object-centric OCEL~\cite{vanderaalstProcessMining3602022}, in which one event can relate to several objects of different types. The prototype supports CSV and XES only.
Three types of process mining are distinguished~\cite{AalstEtAl2012Manifesto}: \emph{Discovery} constructs a process model from a log. Depending on the representational bias of the algorithm, the result is, for example, a directly-follows graph (DFG), a Petri net, a process tree or a BPMN model~\cite{bookProcessMining}. \emph{Conformance checking} compares a log with an existing model to reveal deviations. \emph{Enhancement} extends or repairs an existing model with information from the log, such as waiting times. In the following we will only address discovery. The notation of the discovered model will determine the MM-AR metamodel to which it is exported (Section~\ref{sec:pipeline}).

\subsection{MM-AR Metamodeling Platform}
MM-AR is a web-based, open-source metamodeling platform for spatial conceptual modeling~\cite{muffMMARWebbasedOpensource2026,Fill_2024spatial}. It addresses a limitation of traditional metamodeling platforms, which only support two-dimensional environments~\cite{bookMM-AR}. Like other metamodeling platforms used in enterprise modeling, e.g. ADOxx~\cite{fill_conceptualisation_2013}, it provides a metamodeling client that lets users define their own modeling languages as metamodels, and a modeling client that lets them create models that conform to these metamodels. 

The platform's meta\textsuperscript{2}-model defines the concepts from which all metamodels are built and thereby allows different modeling languages to be used and defined~\cite{muffMMARWebbasedOpensource2026}. Due to spatial conceptual modeling concepts attached to each element, models can then be displayed in both 2D and 3D~\cite{Fill_2024spatial}, either on a 2D screen or using WebXR-enabled devices such as AR/VR headsets or smart glasses. The meta\textsuperscript{2}-model consists of a meta layer and an instance layer. The concepts relevant to this paper are \emph{scene types}, \emph{classes}, \emph{relationclasses} and \emph{attributes} on the meta layer, together with their counterparts on the instance layer. A model is created by instantiating the metaobjects of its metamodel, and each relationclass instance is connected to its source and target elements through role instances. Every object is identified by a universally unique identifier (UUID). The visual appearance of metaobjects is specified with \emph{VizRep}, a domain-specific language for visual representations~\cite{muffMMARWebbasedOpensource2026}.

Technically, the platform is implemented in TypeScript and uses PostgreSQL for persistence~\cite{muffMMARWebbasedOpensource2026}. It consists of several components, each maintained in its own Git repository. A shared data structure module implements the meta\textsuperscript{2}-model as TypeScript classes and ensures a consistent structure across all components. An API server exposes REST endpoints to create, update and delete metamodels, models and their parts, and handles user authentication. The modeling client offers a browser-based interface for viewing, creating and editing models. A separate Docker installation repository bundles all components for deployment and development and initializes a set of predefined metamodels, including Petri nets and BPMN. The prototype presented in this paper builds on the REST API, the modeling client and the Docker setup, and operates directly on model instances that conform to the meta\textsuperscript{2}-model.

\section{Related Work}
\label{sec:related-work}
Three strands of work are related to our approach: the visualization of process mining results, the representation of process and enterprise models in extended reality (XR), and approaches that bring process mining results into immersive environments.

Process discovery results are commonly presented as two-dimensional node-link diagrams, such as directly-follows graphs (DFGs), Petri nets or BPMN models~\cite{bookProcessMining}. This can be achieved today by many open source tools such as ProM~\cite{vandongenProMFrameworkNew2005} or PM4Py~\cite{PM4Py-GPU} or commercial tools. In addition, Yeshchenko and Mendling review event sequence visualization approaches from information visualization and process mining and argue that both fields have developed largely independently~\cite{yeshchenkoSurveyApproachesEvent2024}. In these approaches, the results are however rather visualizations than models, which cannot be further edited in a modeling environment.

Several approaches have been proposed for combining conceptual modeling with virtual and augmented reality~\cite{muffAchievementsFutureOpportunities2023a}.   For example, Oberhauser et al.\ presented VR-BPMN for visualizing, navigating and annotating BPMN models in virtual reality (VR) and compared it empirically with other depiction modes~\cite{oberhauserVRBPMNVisualizingBPMN2018}. The approach was later extended to enterprise architecture models in ArchiMate and BPMN~\cite{oberhauserVREAVirtualReality2019}. However, most of these solutions are dedicated VR applications for specific notations and do not follow a metamodeling based approach as offered by MM-AR.

We could identify two works that use process mining results for virtual reality. Wetzel and Koschmider present a VR environment whose requirements were derived from commercial 2D process mining tools and related work~\cite{wetzelEntwicklungVRUmgebungZur2022}. Process data can be loaded dynamically from source systems, with Celonis used as an example, and explored as a three-dimensional process model based on a DFG. The architecture was designed for extensibility towards other tools. Oberhauser presents VR-ProcessMine, which uses PM4Py as its process mining back end~\cite{oberhauserVRProcessMineImmersiveProcess2022}. It shows a DFG on an initial plate and each process variant on a separate vertical plate, which supports the comparison of variants and illustrates one way of using the third dimension. The author further proposes combining the results with BPMN models displayed in VR-BPMN.
Both environments provide dedicated immersive interaction concepts that the prototype presented here does not offer apart from the ability to display models in VR and AR. They are, however, purpose-built VR clients, their representations are based on DFGs, and the mined results exist only as visualizations within the respective client.

To the best of our knowledge, no existing work transforms process mining results into model instances of a metamodeling platform for XR. We position our approach as complementary to the approaches of Wetzel/Koschmider and Oberhauser~\cite{wetzelEntwicklungVRUmgebungZur2022,oberhauserVRProcessMineImmersiveProcess2022}. Instead of building a dedicated VR client, it maps discovery results onto the existing Petri net and BPMN metamodels of MM-AR, so that mined models can be viewed and edited in 2D and 3D like any manually created model. From an enterprise modeling perspective, discovered models thereby become regular artifacts of the modeling environment that can be refined and reused. The prototype does not yet use the third dimension to encode process information. However, it provides the infrastructure on which such encodings can be implemented and studied in the future.

\section{Tool Selection}
\label{sec:tool-selection}
Among the many available process mining tools, five widely used candidates, selected from the list of processmining.org\footnote{\url{https://processmining.org/software.html}}, were considered: ProM, PM4Py, Apromore, Celonis and Disco. Two criteria guided our selection: an open-source license, in line with MM-AR~\cite{muffMMARWebbasedOpensource2026}, and the possibility of embedding the tool in a separate service without a graphical interface. Commercial tools such as Celonis~\cite{vogelgesangCelonisPQLQuery2022} and Disco~\cite{guntherDiscoDiscoverYour} were therefore excluded, as were free academic licenses, whose conditions may change over time. Apromore was also excluded: its open-source edition, Apromore Core\footnote{\url{https://github.com/apromore/ApromoreCore}}, is no longer actively maintained, while development continues in the commercial edition. Moreover, Apromore is a complete web platform rather than a library, so integrating it would duplicate functionality already provided by MM-AR. Of the remaining tools, ProM~\cite{vandongenProMFrameworkNew2005} is a Java application designed mainly for interactive use, whereas PM4Py~\cite{bertiPM4PyProcessMining2023} is a Python library that can be imported directly into a service, provides discovery algorithms with Petri net and BPMN output, and already serves as the back end of VR-ProcessMine~\cite{oberhauserVRProcessMineImmersiveProcess2022}. We therefore selected PM4Py for our approach.

\section{Architecture}
\label{sec:architecture}
The architecture of our approach addresses the four requirements stated in Section~\ref{sec:introduction}. Process mining runs in a separate Python service rather than inside the existing platform. This turns the language boundary between the TypeScript-based platform and the Python-based process mining ecosystem into a loosely-coupled service boundary. It also leaves the MM-AR server, the shared data structure module and the metamodeling client unchanged (R3), and the service runs as its own container in the Docker installation repository, where it can be maintained or replaced independently. The cost is an additional service call and a separate authentication per request, whose effect on runtime is analyzed in Section~\ref{sec:evaluation}.

The service has two responsibilities: running discovery algorithms on an uploaded event log (R1) and transforming their results into model instances (R2). The second responsibility is model transformation. PM4Py represents discovered models as library-specific Python objects (Petri nets or BPMN diagrams), whereas MM-AR represents models as instances of metamodels that conform to its meta\textsuperscript{2}-model. The platform already provides Petri net and BPMN metamodels whose classes, relationclasses and attributes cover the central concepts of both notations, so the transformation targets these existing metamodels and no new metamodel is required. As a result, mined models become ordinary model instances that can be viewed, edited and displayed in 2D and 3D like manually created ones. Performing the transformation in the service rather than in the client keeps all process-mining-specific logic in one component. The service stores the resulting models through the existing REST API of the MM-AR server and never accesses the database directly.

The user interface is an additive extension of the existing modeling client, so users start process mining from the environment in which they already model. This is intended to lower the entry barrier for non-technical users (R4), but it has not been evaluated empirically yet (Section~\ref{sec:limitations}).

\section{Implementation}
\label{sec:implementation}
The prototype spans three repositories: a new Python repository for the process mining service (pm-service)\footnote{\url{https://github.com/shuerzeler/mmar-pm-service}} and forks of two existing MM-AR repositories. The fork of the modeling client\footnote{\url{https://github.com/shuerzeler/mmar-modeling-client}} adds the user interface for process mining, and the fork of the Docker installation repository\footnote{\url{https://github.com/shuerzeler/mmar-docker-installation}} adds the pm-service as an additional container. The API server, the shared data structure module and the database schema are used without modification. A snapshot of the code is archived at \url{https://doi.org/10.5281/zenodo.21526086}.

\subsection{Service Structure}
The pm-service is implemented with FastAPI. This framework was chosen for three reasons: as a Python framework it allows PM4Py to be called directly within the service, it supports file uploads with request validation, and it can generate an OpenAPI description of the service automatically, which documents the interface for further clients. The modeling client calls a single route, \texttt{/run}, which receives the event log file, the selected discovery algorithm and, for CSV files, the column mapping. Because the modeling client is served from a different port, the service configures cross-origin resource sharing (CORS) so that the browser permits these requests. The code is organised into three folders: \texttt{algorithms} contains one module per discovery algorithm, selected according to the algorithm parameter; \texttt{exporters} contains the Petri net and BPMN exporters; and \texttt{helpers} contains shared functionality. With this separation, a further discovery algorithm or output notation can be added as a new module without changing the route or the existing exporters.

\subsection{Processing Pipeline}
\label{sec:pipeline}
The process mining functionality is accessed from the modeling client. A button in the header menu opens a dialog in which the user uploads an event log, maps its columns (for CSV files only) and selects a discovery algorithm. Submitting the dialog sends a request to the \texttt{/run} route of the pm-service. The dialog then displays either a success message or the error returned by the service. The service processes each request in four steps.
The service first checks the file type (CSV or XES) and, for CSV files, whether the case ID, activity and timestamp columns have been mapped. This validation is kept in the pm-service so that the modeling client does not depend on process-mining-specific logic. The file is then parsed into PM4Py's event log representation. 

Three discovery algorithms provided by PM4Py~\cite{bertiPM4PyProcessMining2023} are available: the Alpha Miner~\cite{vanderaalstWorkflowMiningDiscovering2004}, the Heuristics Miner~\cite{weijtersProcessMiningHeuristicsMiner2006,weijtersFlexibleHeuristicsMiner2011} and the Inductive Miner~\cite{leemansDiscoveringBlockStructuredProcess2013}.
All three produce a Petri net, and the Inductive Miner can alternatively produce a BPMN model. The Alpha Miner serves as a well-known baseline, the Heuristics Miner is designed to cope with infrequent behaviour, and the Inductive Miner guarantees sound, block-structured models. 

The resulting PM4Py object is mapped to a model instance of the existing Petri net or BPMN metamodel of MM-AR. The exporters retrieve the UUIDs of the required metamodel, classes, relationclasses and attributes, so that no identifiers are hard-coded and the exporters work on any MM-AR installation that contains these metamodels. PM4Py provides the structure of a model but no layout, whereas the MM-AR meta\textsuperscript{2}-model requires explicit coordinates for every element. The exporters therefore build a graph of the model with NetworkX~\cite{hagbergExploringNetworkStructure2008}, compute a layout with Graphviz~\cite{ellsonGraphvizOpenSource2002} and scale the resulting coordinates to the MM-AR canvas. 

The complete model instance is sent to the API server in a single request. Its name is prefixed with a timestamp, which distinguishes mined models from manually created ones and avoids name clashes. In the current version, the pm-service authenticates with its own credentials instead of reusing the session token of the modeling client (see Section~\ref{sec:limitations}).

\section{Demonstration and Evaluation}
\label{sec:evaluation}
In the following we first demonstrate the prototype as a proof-of-concept and then evaluate it technically. The technical evaluation addresses two questions: (EQ1) Do the MM-AR model instances created by the exporters structurally match the models produced by PM4Py? (EQ2) How long does an end-to-end run take, and how does the runtime develop with the size of the event log? Section~\ref{sec:datasets} introduces the event logs used throughout. Section~\ref{sec:demonstration} walks through the process mining pipeline end to end, including the prototype's behaviour under invalid input, and Section~\ref{sec:evaluation} answers EQ1 and EQ2. The quality of the discovered models is not assessed, since it is determined by PM4Py rather than by the integration. An empirical evaluation of how the spatial representation affects process comprehension, for example through a user study, is beyond the scope of this paper (see Sections~\ref{sec:limitations} and~\ref{sec:conclusion}).

\subsection{Event logs}
\label{sec:datasets}
Four publicly available event logs were selected so that they differ in size, origin and file format (see Table~\ref{tab:logs}). The small logs allow a visual comparison between the MM-AR and PM4Py outputs, the range of sizes allows runtime behaviour to be observed, and the CSV log exercises the column mapping in the dialog.

The \emph{Running Example} (RE, \texttt{running-example.xes})\footnote{\url{https://github.com/process-intelligence-solutions/pm4py/blob/release/tests/input_data/running-example.xes}} is a small artificial log commonly used to demonstrate PM4Py.
The \emph{PDC 2024} log (\texttt{pdc2024\_000000.xes})\footnote{\url{https://data.4tu.nl/datasets/3cfcdbb7-c909-4f60-8bec-62c780598047/1}} is the first test log of the Process Discovery Contest 2024, which compared discovery algorithms on artificially generated event logs at the International Conference on Process Mining 2024~\cite{delgadoProcessMiningWorkshops2025}.
The \emph{Road Traffic Fine Management} log (RTFM)\footnote{\url{https://data.4tu.nl/articles/dataset/Road_Traffic_Fine_Management_Process/12683249}} is a large real-life log with 150'370 traces that has previously been used in the evaluation of PM4Py-GPU~\cite{PM4Py-GPU}.
The \emph{P2P} log (\texttt{P2P.csv})\footnote{\url{https://processmind.com/resources/docs/example-files/log-files}} is a medium-sized example log provided by Processmind.
All logs were used as downloaded, without preprocessing. For P2P, the case ID, activity and timestamp columns only had to be mapped in the dialog.

\begin{table}[tb]
 \centering\footnotesize
 \caption{Event logs used in the evaluation}
 \label{tab:logs}
 \begin{tabular}{@{}llrr@{}}
   \toprule
   \textbf{Log} & \textbf{Format, origin} & \textbf{Cases} & \textbf{Events} \\
   \midrule
   Running Example & XES, artificial & 6 & 42 \\
   PDC 2024 & XES, artificial & 1{'}000 & 14{'}924 \\
   Road Traffic Fine Management & XES, real-life & 150{'}370 & 561{'}470 \\
   P2P & CSV, artificial & 934 & 7{'}782 \\
   \bottomrule
 \end{tabular}
\end{table}
\subsection{Demonstration of the Prototype}
\label{sec:demonstration}
The walkthrough uses the P2P.csv log (Section~\ref{sec:datasets}) and the Alpha Miner. A dialog added to the modeling client lets the user upload an event log, map the case ID, activity and timestamp columns (for CSV files only) and select a discovery algorithm. The file and the selected options are then sent to the process mining service. After a successful run, the page must be reloaded manually to display the mined model.

\begin{figure}[htbp]
  \centering
  \includegraphics[width=\textwidth]{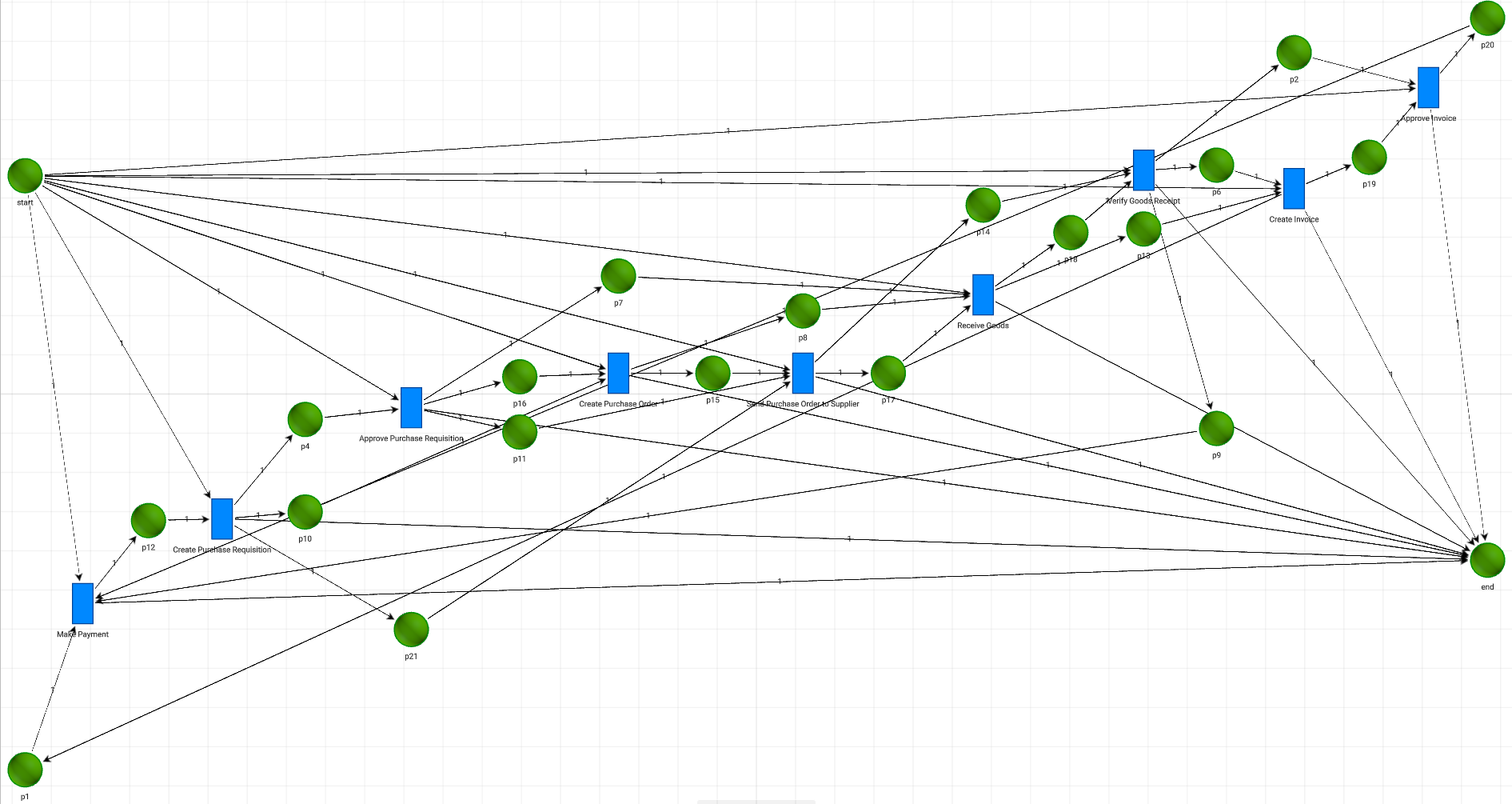}
  \caption{Petri net mined from P2P.csv with the Alpha Miner, shown in 2D mode}
  \label{fig:d6-petrinet}
\end{figure}

After the reload, the mined Petri net appears as a regular model instance of the platform's existing Petri net metamodel (Figure~\ref{fig:d6-petrinet}). It can be edited in the same way as a manually created model: places and transitions can be repositioned by dragging them on the canvas, their names can be changed, the number of tokens of a place can be set in the properties pane on the right and arc weights can be adjusted. The source place holds one token, because the exporter adopts the initial marking returned by PM4Py.

\begin{figure}[htbp]
  \centering
  \includegraphics[width=0.6\textwidth]{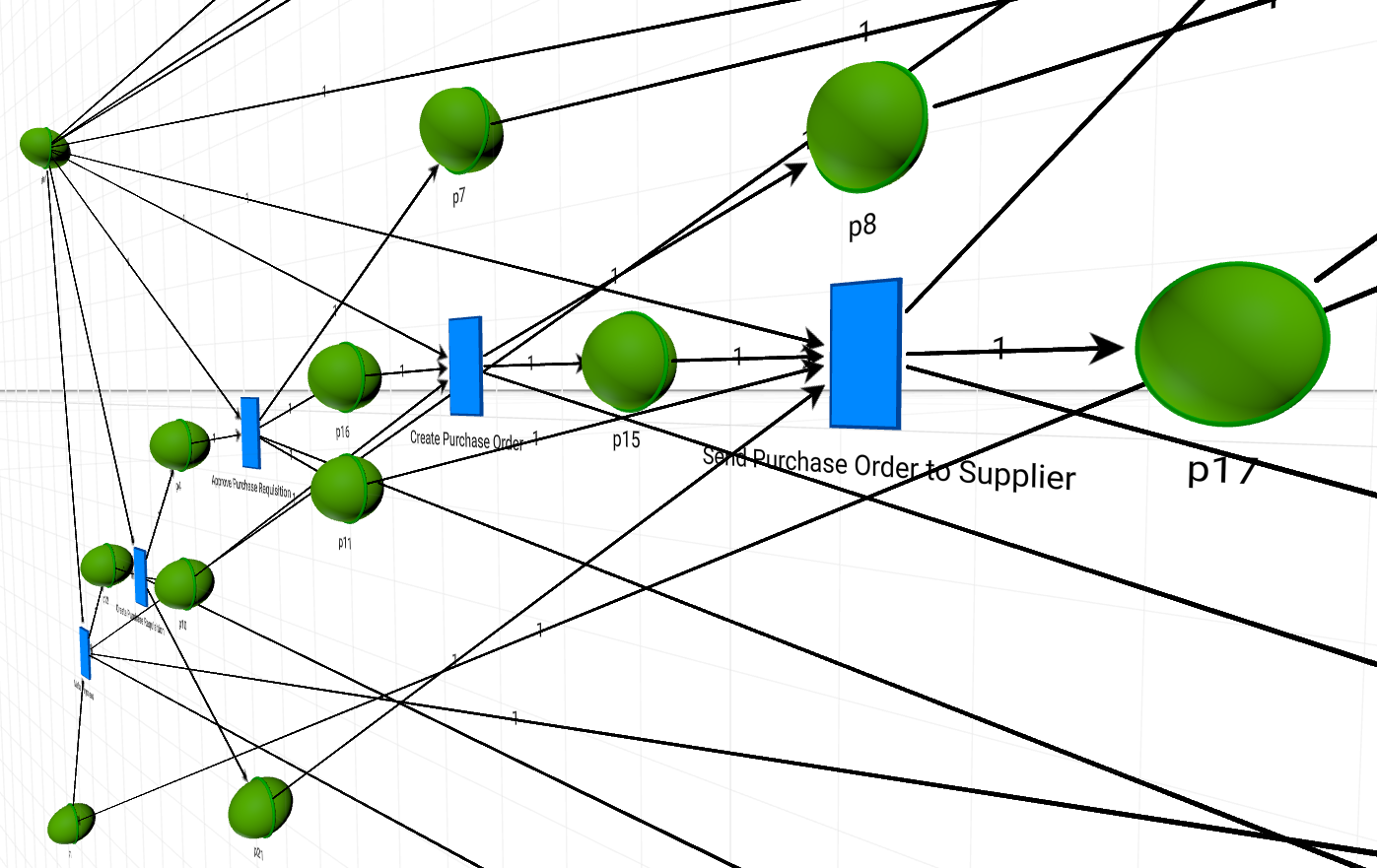}
  \caption{Detail of the Petri net from Figure~\ref{fig:d6-petrinet} in 3D mode}
  \label{fig:d7-petrinet-3d}
\end{figure}

The same model can also be displayed in the platform's 3D mode (Figure~\ref{fig:d7-petrinet-3d}).

Other algorithms produce different models from the same log. The Inductive Miner, for example, discovers silent transitions, which are labelled \texttt{tau} followed by a generated identifier. The Inductive Miner can also produce a BPMN model, using a separate PM4Py function.

Invalid input and connection problems are reported in the dialog: unsupported file formats (e.g.\ .xlsx) are rejected with a list of the accepted formats, runs with incomplete column mappings cannot be started, and an unreachable process mining service results in the browser message ``Failed to fetch''.

\subsection{Evaluation of the Prototype}

The technical evaluation addresses two questions. \textbf{EQ1 (structural correctness):} does the MM-AR model instance created by the exporters contain the same elements and connections as the model returned by PM4Py? \textbf{EQ2 (runtime overhead):} how much time does the integration add to the PM4Py discovery step, and how does this additional time relate to the size of the event log and of the discovered model? Since PM4Py performs the discovery, the quality of the discovered models (e.g.\ fitness or precision) is not assessed. The evaluation targets only the parts of the pipeline contributed by this work.

A runtime comparison with other tools such as Apromore or Celonis is not performed. These tools run as cloud services, whereas the prototype runs locally, so measured differences would mainly reflect hardware and deployment rather than the integration itself. Instead, the PM4Py runtime measured on the same machine serves as the baseline from which the overhead of the integration is derived.

Both questions are answered with an evaluation script included in the prototype repository under evaluation/evaluation.py. For each of the four event logs (Section~\ref{sec:datasets}) and each discovery option, the script runs PM4Py directly, stores the PM4Py visualization in a results folder, calls the pm-service and compares the resulting models.

\begin{figure}[tb]
  \centering
  \includegraphics[width=\textwidth]{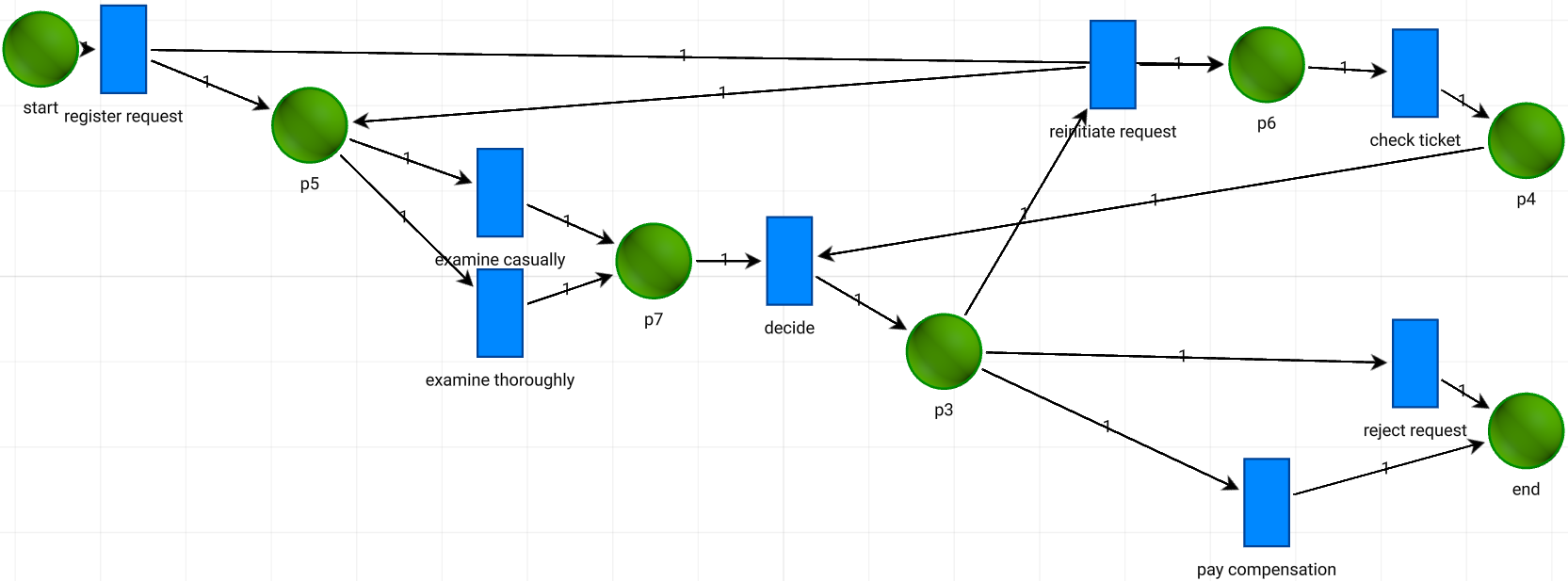}
  
  \includegraphics[width=\textwidth]{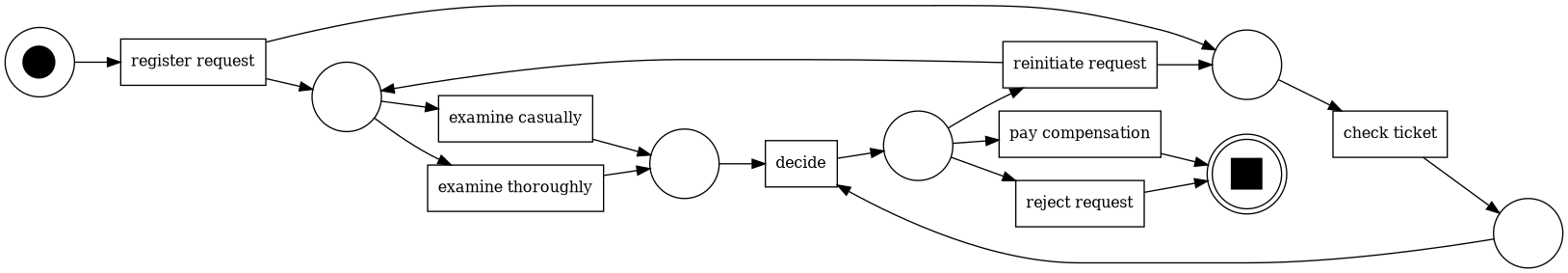}
  \caption{Petri net discovered with the Alpha Miner from running-example.xes: MM-AR model (top) and PM4Py visualisation (bottom)}
  \label{fig:comparison-alpha}
\end{figure}

Structural correctness is checked in two steps. First, the MM-AR model is compared visually with the PM4Py visualization. As this is only feasible for small models, the running-example.xes log is used. Figure~\ref{fig:comparison-alpha} shows the Petri net discovered with the Alpha Miner, with the MM-AR model at the top and the PM4Py visualization at the bottom. The layouts differ slightly: in the MM-AR model, one arc passes through a transition and the end place is positioned differently. However, every arc connects the same place and transition in both models.

\begin{figure}[tb]
  \centering
  \includegraphics[width=\textwidth]{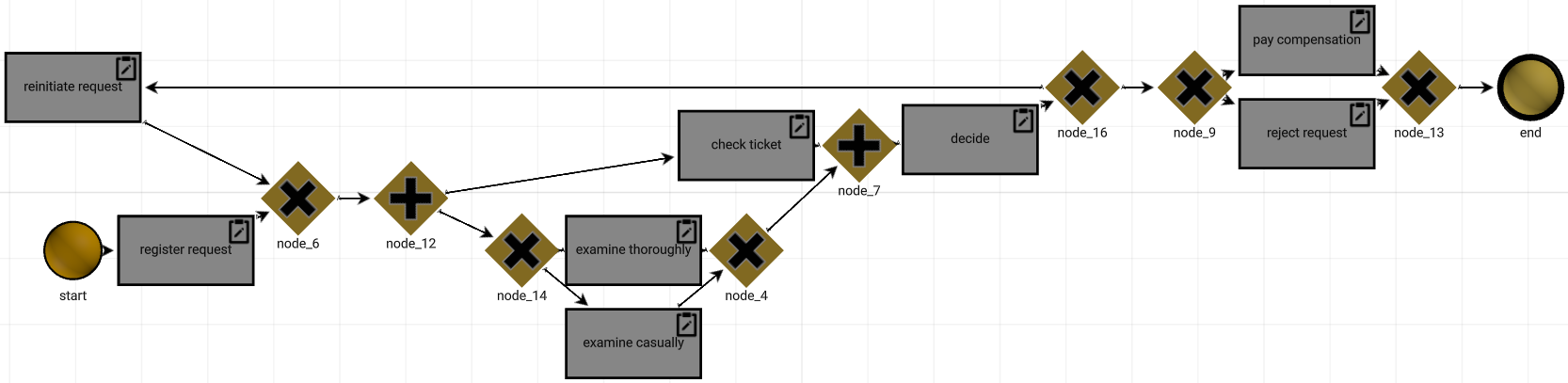}
  \includegraphics[width=\textwidth]{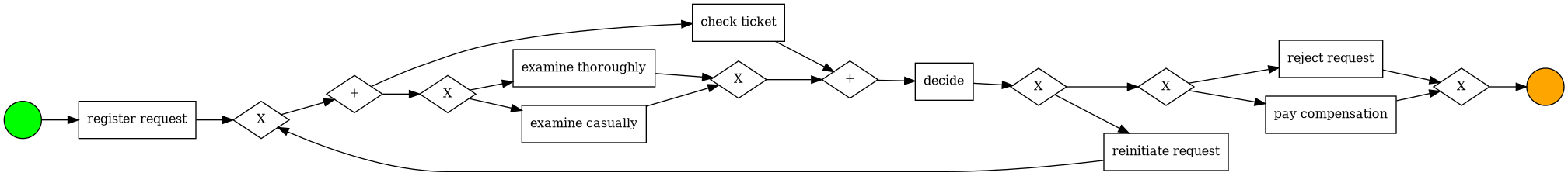}
  \caption{BPMN model discovered with the Inductive Miner from running-example.xes: MM-AR model (top) and PM4Py visualisation (bottom)}
  \label{fig:comparison-bpmn-inductive}
\end{figure}

Figure~\ref{fig:comparison-bpmn-inductive} shows the corresponding comparison for the BPMN model discovered with the Inductive Miner. Tasks, events and gateways are connected identically in both models. The only differences concern rendering: start and end events are drawn differently, and sequence flows are straight in MM-AR but curved in PM4Py. The same visual comparison was carried out for the two remaining discovery options (Inductive Miner and Heuristics Miner, both producing Petri nets) and showed no structural differences; these comparisons are not shown for reasons of space but can be accessed at~\cite{hurzelerExtendingMMARPlatform2026}.

Visual comparison does not scale to the larger logs. Therefore, as a second step, the evaluation script compares the number of elements in the PM4Py model with the number of elements in the corresponding MM-AR model instance: places, transitions and arcs for Petri nets, and nodes (tasks, events and gateways) and sequence flows for BPMN models. The counts match for all discovery options and event logs (Table~\ref{tab:structural-correctness-results}).

Equal counts are a necessary but not a sufficient condition for structural equivalence, since they do not show that elements are connected or labelled identically. A full equivalence check, which would have to map silent transitions and duplicate labels between the two representations, was not implemented. Taken together, the visual comparison on the small log and the matching counts on all four logs are consistent with a correct transformation, but do not prove it for the larger models.

\begin{table}
  \centering
  \footnotesize
  \setlength{\tabcolsep}{2.5pt}
\caption{Structural correctness of the discovered models. Each cell gives the number of elements in the MM-AR model instance / in the PM4Py model. P = places, T = transitions, A = arcs, N = nodes, F = flows. A tick in the Match column means that all counts in the row are identical.}
\label{tab:structural-correctness-results}
  \begin{tabular}{@{}l ccc ccc ccc cc c@{}}
    \toprule
    & \multicolumn{9}{c}{\textbf{Petri net}} & \multicolumn{2}{c}{\textbf{BPMN}} & \\
    \cmidrule(lr){2-10} \cmidrule(lr){11-12}
    & \multicolumn{3}{c}{\textbf{Alpha}} & \multicolumn{3}{c}{\textbf{Inductive}} & \multicolumn{3}{c}{\textbf{Heuristic}} & \multicolumn{2}{c}{\textbf{Inductive}} & \\
    \cmidrule(lr){2-4} \cmidrule(lr){5-7} \cmidrule(lr){8-10} \cmidrule(lr){11-12}
    \textbf{Dataset} & P & T & A & P & T & A & P & T & A & N & F & \textbf{Match} \\
    \midrule
    Running Example    & 7/7   & 8/8   & 19/19 & 9/9   & 10/10 & 22/22   & 13/13 & 15/15   & 35/35   & 18/18   & 21/21   & \checkmark \\
    PDC 2024           & 28/28 & 25/25 & 67/67 & 57/57 & 87/87 & 192/192 & 68/68 & 106/106 & 259/259 & 105/105 & 153/153 & \checkmark \\
    RTFM & 12/12 & 11/11 & 29/29 & 29/29 & 34/34 & 82/82   & 25/25 & 40/40   & 92/92   & 49/49   & 68/68   & \checkmark \\
    P2P                & 21/21 & 9/9   & 56/56 & 4/4   & 12/12 & 24/24   & 29/29 & 40/40   & 90/90   & 15/15   & 23/23   & \checkmark \\
    \bottomrule
  \end{tabular}
\end{table}
Table~\ref{tab:average-runtime-results} reports the mean runtime over three runs of the evaluation script for each dataset and algorithm. For each configuration, the PM4Py pipeline time (event log loading, discovery and Graphviz visualisation) and the end-to-end time of the corresponding pm-service call were measured; their difference is reported as the MM-AR overhead. All runs were executed on the same machine using the forked Docker installation repository. Absolute times therefore depend on this setup and hardware.

\begin{table}
  \centering
  \footnotesize
  \setlength{\tabcolsep}{3pt}
\caption{Mean runtime in seconds over three runs, by dataset and algorithm. PM4Py = PM4Py pipeline (event log loading, discovery and Graphviz visualisation); Total = end-to-end time of the pm-service call; OH = MM-AR overhead (Total $-$ PM4Py).}
  \label{tab:average-runtime-results}
  \begin{tabular}{@{}l@{\hspace{8pt}}rrr@{\hspace{10pt}}rrr@{\hspace{10pt}}rrr@{\hspace{10pt}}rrr@{}}
    \toprule
    & \multicolumn{9}{c}{\textbf{Petri net}} & \multicolumn{3}{c}{\textbf{BPMN}} \\
    \cmidrule(lr){2-10} \cmidrule(l){11-13}
    & \multicolumn{3}{c}{\textbf{Alpha}} & \multicolumn{3}{c}{\textbf{Inductive}} & \multicolumn{3}{c}{\textbf{Heuristic}} & \multicolumn{3}{c}{\textbf{Inductive}} \\
    \cmidrule(lr){2-4} \cmidrule(lr){5-7} \cmidrule(lr){8-10} \cmidrule(l){11-13}
    \textbf{Dataset} & PM4Py & Total & OH & PM4Py & Total & OH & PM4Py & Total & OH & PM4Py & Total & OH \\
    \midrule
    Running Example    &  0.19 &  8.71 &  8.52 &  0.07 &  8.53 &  8.46 &  0.12 & 11.47 & 11.36 &  0.06 &  8.19 &  8.13 \\
    PDC 2024           &  0.33 & 17.63 & 17.30 &  2.27 & 43.44 & 41.18 &  1.29 & 54.18 & 52.89 &  2.23 & 32.52  & 30.29 \\
    RTFM & 27.48 & 36.60 &  9.12 & 29.41 & 47.06 & 17.65 & 30.70 & 50.02 & 19.32 & 28.28 & 42.19 & 13.91 \\
    P2P                &  0.54 & 15.06 & 14.52 &  0.22 &  8.56 &  8.35 &  0.57 & 22.59 & 22.01 &  0.20 &  8.36 &  8.16 \\
    \bottomrule
  \end{tabular}
\end{table}
The overhead of the integration is not constant. It ranges from about 8\,s for the small models of the Running Example to about 53\,s for the Heuristics Miner on the PDC 2024 log. Comparing Table~\ref{tab:average-runtime-results} with Table~\ref{tab:structural-correctness-results} shows that the overhead grows with the number of elements of the discovered model rather than with the size of the event log: across the 16 configurations, overhead and model size are strongly correlated (Pearson's $r \approx 0.99$). For the Road Traffic Fine Management log, the largest log, PM4Py dominates the runtime (means of 27 to 31\,s), while the overhead (9 to 19\,s) stays below that of the smaller PDC 2024 log, whose discovered models are larger. Accordingly, the Heuristics Miner, which yields the models with the most elements on all four logs, has the highest total runtime. The source of the size-dependent overhead was not profiled.

These results are based on four logs, only one of which is large, and on three runs per configuration. The individual runs are consistent: for each configuration, the overheads of the three runs differ by at most 4.4\,s (PDC 2024, Heuristics Miner: 50.5 to 54.9\,s), and the standard deviation is at most 2.3\,s. Relative to the mean, the variation is largest where the overhead is small, namely for the Alpha Miner on RTFM and the Heuristics Miner on the Running Example (coefficient of variation of about 15\,\

In summary, the evaluation targets what the prototype adds, i.e. the conversion of PM4Py results into MM-AR model instances and its runtime cost, rather than the quality of the discovery algorithms, which depends on PM4Py. For all four logs and all four configurations, the element counts of the MM-AR model instances match those of the PM4Py models. Together with the visual comparison for the running example, this supports the structural correctness of the exporters. The integration adds an overhead of about 8 to 53\,s per call, which grows with the size of the discovered model.

\section{Limitations and Threats to Validity}
\label{sec:limitations}

On the technical side, the process mining service authenticates independently instead of reusing the client's session token, which may contribute to the overhead of about 8\,s that occurs even for the smallest models (lowest single run: 7.8\,s), although the individual steps were not measured separately. The overhead also appears to grow with the size of the mined model rather than of the event log: it peaks at 52.89\,s for the largest model (Heuristics Miner, PDC 2024), while the much larger Road Traffic Fine Management log stays below 20\,s. Functionally, the prototype supports only CSV and XES logs, process discovery with three algorithms at default parameters, and BPMN output only for the Inductive Miner. Object-centric logs, conformance checking and quality measures are not supported, although PM4Py provides them~\cite{bertiPM4PyProcessMining2023}. After mining, the page must be reloaded manually.

The most important limitation is conceptual. Mined models can be opened in the MM-AR 3D environment, but their coordinates come from a scaled 2D Graphviz layout, so the third dimension carries no information.

Structural correctness was checked by element counts, with a visual comparison only for the smallest log, and runtimes are averages of three runs on one machine. The scalability statement therefore rests on a single large log. The four clean, public logs include three artificial and one real-life log, so results may not transfer to noisy real-life data. Usability for non-technical users was not evaluated with users. Finally, while the decoupled architecture is likely transferable, the exporters are specific to PM4Py and the MM-AR meta\textsuperscript{2}-model.

\section{Conclusion and Future Work}
\label{sec:conclusion}
This paper addressed the missing integration of process mining into a modeling and metamodeling platform for extended reality. As a proof of concept, the MM-AR platform was extended with a separate Python service based on PM4Py, deployed as an additional container in the existing Docker installation repository and connected to the platform via REST, together with an extension of the modeling client. Users can upload an event log, map its columns and select a discovery algorithm. The results are stored as model instances conforming to the existing Petri net and BPMN metamodels, so discovered models can be viewed and edited in the 2D and 3D environments of the platform in the same way as manually created models. From an enterprise modeling perspective, the integration turns process mining output into metamodel-conformant model artefacts rather than static visualisations, without changes to the platform core.

The technical evaluation supports the feasibility of this integration within its scope. For all four event logs and all implemented discovery options, the element counts of the MM-AR models matched those of the PM4Py models, and a visual comparison for the smallest log showed no missing elements. 
The most important open question is whether spatial encoding improves the comprehension of process models. The prototype displays mined models in 3D but does not yet assign meaning to the third dimension. Future work should therefore design a 3D layout for mined models, for example by placing process variants on separate planes as proposed by Oberhauser~\cite{oberhauserVRProcessMineImmersiveProcess2022}, and evaluate it in a user study that compares comprehension between the 2D and 3D modes of the platform. It could be explored how the use of AI can help in this context to support users in 3D interaction~\cite{fillCMAGFrameworkConceptual}. 
On the technical side, further extensions can be the configuration of algorithm parameters, such as the noise threshold of the Heuristics Miner, the display of quality measures for mined models, and support for object-centric process mining. The latter is of particular interest because the third dimension could be used to separate multiple object types, which is difficult in 2D notations. 
In augmented reality, a discovered model could be placed next to the physical equipment whose process it describes, which may help users relate the model to its environment.

\section*{Declaration on Generative AI}
For this paper the authors used Anthropic's Claude to paraphrase and reword in order to precisely transport ideas, for grammar and spelling check, for formatting assistance and to improve writing style. All content was reviewed and edited by the authors. The authors take full responsibility for the content of this paper.

\bibliography{references_Daniel}

\end{document}